\documentclass[aps,prb,twocolumn,showpacs,amsmath,amssymb]{revtex4}
\usepackage{amssymb, amsmath, amsbsy}
\usepackage{pdflscape}
\usepackage{multirow}
\usepackage{hhline}
\usepackage{graphicx}
\usepackage{subfigure}
\usepackage{times}
\usepackage{doi}

\begin{document}

\title{Vortex states and Majorana fermions in spin-orbit coupled semiconductor-superconductor hybrid structures}
\author{Kristofer Bj\"{o}rnson}
\affiliation{Department of Physics and Astronomy, Uppsala University, Box 516, S-751 20 Uppsala, Sweden}
\author{Annica M. Black-Schaffer}
\affiliation{Department of Physics and Astronomy, Uppsala University, Box 516, S-751 20 Uppsala, Sweden}
\date{\today}

\begin{abstract}
We study the energy spectrum of a vortex core in a two-dimensional semiconductor with Rashba spin-orbit interaction and proximity-coupled to a conventional superconductor and a ferromagnetic insulator. We perform self-consistent calculations using the microscopic tight-binding Bogoliubov-de Gennes method on a lattice and confirm the existence of Majorana fermions in the non-trivial topological phase. We also find two different topologically trivial bulk superconducting phases, only differing in the type of vortex core structure they support and separated by a zero-energy excitation. Furthermore, we find an asymmetry in the energy spectrum with respect to both Zeeman splitting and vortex rotation direction and explain its physical origin.
\end{abstract}
\pacs{74.90.+n, 03.65.Vf, 03.67.Lx, 74.45.+c}

\maketitle

\section{Introduction}
In condensed matter physics the topic of topology has a rich history. For several decades topology has played an important role in the study of phenomena such as nematics, domain walls, and vortices, through the winding numbers of the corresponding order parameters.\cite{RevModPhys.51.591, TopologicalQuantumNumbersInNonrelativisticPhysics} With the  discovery of the quantum Hall effect\cite{PhysRevLett.45.494, PhysRevB.23.5632, PhysRevLett.49.405} a second type of topological theory emerged, in which topology in reciprocal rather than real space is important.\cite{TopologicalQuantumNumbersInNonrelativisticPhysics} Inspired by the topological concepts introduced through the quantum Hall effect, the existence of a new class of materials called topological insulators was recently theoretically predicted\cite{PhysRevLett.95.226801, PhysRevLett.95.146802, PhysRevLett.96.106802, Science.1133734} and subsequently experimentally verified.\cite{Science.1148047} This has opened up a large research field, which also includes the very closely related concept of topological superconductors.\cite{RevModPhys.82.3045, RevModPhys.83.1057} 

On an initially unrelated note, Majorana introduced in 1937 a slightly modified version of the Dirac equation, where the particles are their own anti-particles.\cite{EttoreMajorana} The neutrino has been proposed to be a Majorana fermion, but experimental investigations has so far failed to confirm the existence of any fundamental fermions of the Majorana type.\cite{nphys1380} More recently, Majorana fermions have been proposed to exist as effective quasi-particle excitations on edges or in vortex cores of either spinless $p+ip$-wave superconductors or fractional quantum Hall systems with filling fraction $\nu = \frac{5}{2}$.\cite{JETPLett.70.609, PhysRevB.61.10267, RepProgPhys.75.076501} 
From a theoretical point of view this provides an interesting playground, where concepts from both real and reciprocal topological theories, in the form of vortices and quantum Hall-like physics, respectively, meets high-energy physics in the long standing quest to discover Majorana fermions.\cite{nphys1380, TheUniverseInAHeliumDroplet} In addition, and application-wise more relevant, the non-Abelian nature of the Majorana fermions makes them compelling candidates for implementing fault-tolerant topological quantum computation through braiding schemes.\cite{Nayak08}

Recently, an effective two-dimensional (2D) spinless $p+ip$-wave superconducting state has been proposed in both topological insulators \cite{PhysRevLett.100.096407} and ferromagnetic Rashba spin-orbit coupled semiconductors  \cite{PhysRevLett.103.020401, PhysRevB.81.125318, PhysRevB.82.134521, PhysRevLett.104.040502, PhysRevLett.105.077001, PhysRevLett.105.177002, RepProgPhys.75.076501} in proximity to a conventional $s$-wave superconductor. The maturity of semiconductor technology, large spin-orbit coupling, \cite{Yang06, Sakano13} and experimentally demonstrated superconducting proximity effect \cite{PhysRevB.55.8457} make especially the hybrid semiconductor-superconductor alternative very promising and possible experimental signatures of Majorana physics has already been reported in such nanowires.\cite{Science.1222360, nphys2429, nphys2479} Although proposals exists for braiding operations in a network of 1D nanowires,\cite{nphys1915} vortices offer a more direct experimental route to braiding, important for both testing the non-Abelian nature of Majorana fermions as well as implementing computational operations.

While several studies based on a continuum model have established the existence of a Majorana mode in the vortex core in ferromagnetic hybrid semiconductor-superconductor structures,\cite{PhysRevLett.104.040502, PhysRevB.82.134521, PhysRevLett.103.020401,PhysRevB.82.174506} no detailed microscopic self-consistent study exists.
In this article we therefore study a microscopic model of a 2D Rashba spin-orbit coupled semiconductor with proximity-induced $s$-wave superconductivity and Zeeman exchange splitting. We perform self-consistent calculations using the microscopic tight-binding Bogoliubov-de Gennes method\cite{Superconductivity_of_metal_and_alloys} on a lattice and obtain spatially well separated Majorana fermion solutions when the parameters are tuned such that the whole system enters the topologically non-trivial phase. In addition to the topologically non-trivial phase we find two topologically trivial bulk superconducting phases, distinguished by their vortex core magnetization. These two phases are separated by a zero-energy vortex core mode, explicitly demonstrating that zero-energy states in these hybrid structures can occur outside the topologically non-trivial phase. Moreover, we find a distinct asymmetry in the vortex core energy spectrum as function of either Zeeman splitting or vortex rotation direction. We show that this asymmetry is also present in the continuum model and provide a physical understanding of its origin. These results establish a remarkably rich behavior of the vortex core energy spectrum, including different zero-energy states. With the energy spectrum being the primary experimental screening tool for identifying Majorana fermions, our results should provide a guideline in the search for vortex Majorana states.

\section{Method}
\begin{figure}
\includegraphics[width=225pt]{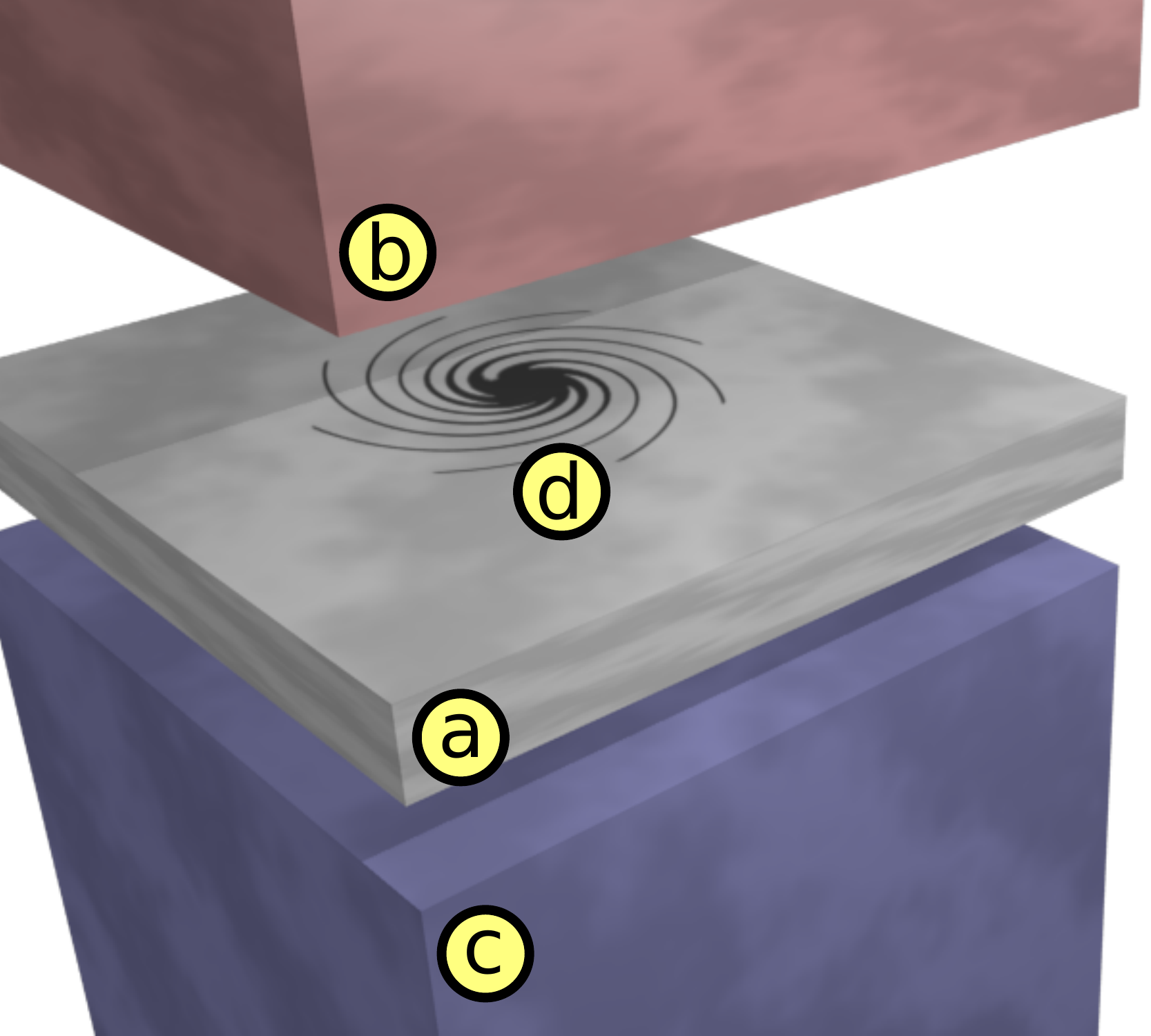}
\caption{(Color online) Schematic setup with (a) thin semiconductor layer with Rashba spin-orbit interaction into which \textit{s}-wave superconductivity and Zeeman splitting are introduced by proximity effect from (b) a conventional superconductor and (c) a ferromagnetic insulator. We study (d) a superconducting vortex in this system.}
\label{Figure:Conceptual_setup}
\end{figure}

We consider a 2D semiconductor on a square lattice with Rashba spin-orbit interaction, spin-singlet \textit{s}-wave superconductivity and Zeeman splitting. This system can be realized in a thin Rashba spin-orbit coupled semiconductor layer sandwiched between a conventional superconductor and a ferromagnetic insulator, which by proximity-effect induce the \textit{s}-wave superconductivity and Zeeman splitting, see Fig.~\ref{Figure:Conceptual_setup}. The Hamiltonian for this system can be written as\cite{PhysRevLett.103.020401,PhysRevB.82.134521, PhysRevB.84.180509}
\begin{align}
\mathcal{H} &= \mathcal{H}_{kin} + \mathcal{H}_{V_z} + \mathcal{H}_{SO} + \mathcal{H}_{sc},
\label{Equation:Tight_binding_Hamiltonian} \\ 
\mathcal{H}_{kin} &= -t\sum_{\langle\mathbf{i},\mathbf{j}\rangle,\sigma}c_{\mathbf{i}\sigma}^{\dagger}c_{\mathbf{j}\sigma} - \mu\sum_{\mathbf{i},\sigma}c_{\mathbf{i}\sigma}^{\dagger}c_{\mathbf{i}\sigma}, \nonumber \\ 
\mathcal{H}_{V_z} &= -V_z\sum_{\mathbf{i},\sigma,\sigma'}(\sigma_z)_{\sigma\sigma'}c_{\mathbf{i}\sigma}^{\dagger}c_{\mathbf{i}\sigma'}, \nonumber \\ 
\mathcal{H}_{SO} &= -\frac{\alpha}{2}\sum_{\mathbf{i}}\left[ (c_{\mathbf{i}-\hat{x}\downarrow}^{\dagger}c_{\mathbf{i}\uparrow} - c_{\mathbf{i}+\hat{x}\downarrow}^{\dagger}c_{\mathbf{i}\uparrow})\right. \nonumber \\ 
&\textrm{\;\;\;\;\;\;\;\;\;\;\;\;\;\;\;\;}\left.+ i(c_{\mathbf{i}-\hat{y}\downarrow}^{\dagger}c_{\mathbf{i}\uparrow} - c_{\mathbf{i+\hat{y}}\downarrow}^{\dagger}c_{\mathbf{i}\uparrow}) + {\rm H.c.}\right], \nonumber \\ 
\mathcal{H}_{sc} &= \sum_{\mathbf{i}}\Delta_{\mathbf{i}}(c_{\mathbf{i}\uparrow}^{\dagger}c_{\mathbf{i}\downarrow}^{\dagger} + H.c.) \nonumber.
\end{align}
Here $\mathbf{i}$ and $\mathbf{j}$ are site indices on the square lattice, $\sigma$ is the spin index, and $c_{\mathbf{i}\sigma}^{\dagger} (c_{\mathbf{i}\sigma})$ is the electronic creation (annihilation) operator. $t$, $\mu$, $V_z$, $\alpha$ and $\Delta$ are the nearest neighbor hopping, chemical potential, Zeeman splitting, Rashba spin-orbit interaction, and superconducting order parameter, respectively. For concreteness, we consider a lightly hole-doped semiconductor by setting $\mu = 4$ and measure all energies relative to the kinetic term, which we set to $t = 1$. We are interested in the properties of superconductor vortices in this system, for which the order parameter can be written as  $\Delta(r,\theta) = |\Delta(r,\theta)|e^{in\theta}$ around some vortex center $r = 0$, where $n = \pm 1$ determines the vortex rotation direction.

We solve Eq.~(\ref{Equation:Tight_binding_Hamiltonian}) self-consistently within the Bogoliubov-de Gennes formulation.\cite{Superconductivity_of_metal_and_alloys} Due to the spin-orbit coupling and Zeeman field, a doubling of the number of degrees of freedom is necessary, resulting in two eigenstates for every ordinary electronic degree of freedom.
The self-consistent calculation is carried out by specifying a vortex-like initial configuration
$\Delta^{(0)} = e^{in\theta}$.
The superconducting pairing potential $V_{sc}$ is then introduced and used in subsequent steps to calculate
\begin{align}
\begin{split}
\Delta_{\mathbf{i}}^{(m+1)} &= -V_{sc}\langle c_{\mathbf{i}\downarrow}c_{\mathbf{i}\uparrow}\rangle^{(m)}\\
&= -V_{sc}\sum_{E_{\nu} < 0} u_{\nu\mathbf{i}\uparrow}^{(m)}v_{\nu\mathbf{i}\downarrow}^{(m)*},
\end{split}
\end{align}
where $u_{\nu\mathbf{i}\uparrow} (v_{\nu\mathbf{i}\downarrow})$ is the electron up (hole down) component on site $\mathbf{i}$ in the $\nu$'th eigenstate and $m$ is the iteration step. $\Delta^{(m+1)}$ is then fed back into the calculation until a well converged $\Delta$ is obtained. It is thus the pairing potential $V_{sc}$ together with vortex rotation direction $n$ that self-consistently determines $\Delta$. If not otherwise stated, calculations presented here are done on a lattice of size $39\times 39$. However, calculations were also preformed for lattice sizes down to as small as $14\times 14$ to ensure that the results are not sensitive to the lattice size. For the smallest lattice sizes interference effects between the vortex and edge states become increasingly important, but no other differences were found with varying lattice size.

The Hamiltonian in Eq.~(\ref{Equation:Tight_binding_Hamiltonian}) has several topologically non-trivial phases. For a lightly hole-doped semiconductor as considered here, the condition for being inside the only experimentally accessible topologically non-trivial phase is\cite{PhysRevB.82.134521}
\begin{equation}
|\Delta| < |V_z| < \sqrt{\mu^2 + |\Delta|^2}.
\label{Equation:Topologically_non_trivial_phase_condition_1}
\end{equation}
However, the superconducting order parameter is determined self-consistently and it will respond to changes in $V_z$ and approach zero for moderately to large $V_z$. At this point the superconducting gap is destroyed and the topological band theory breaks down. As this will happen long before the upper bound in Eq. (\ref{Equation:Topologically_non_trivial_phase_condition_1}) is violated, the condition for the non-trivial phase can be written as
\begin{equation}
0 < |\Delta| < |V_z|.
\label{Equation:Topologically_non_trivial_phase_condition_2}
\end{equation}

Having fixed $t = 1$, $\mu = 4$, and replaced $\Delta$ with $V_{sc}$ and $n$ as external parameters there are four remaining parameters in the Hamiltonian in Eq.~(\ref{Equation:Tight_binding_Hamiltonian}). These are the strength of the Zeeman splitting ($V_z$), Rashba spin-orbit interaction ($\alpha$), superconducting pairing potential ($V_{sc}$) and vortex rotation direction ($n$).

\section{Results and discussion}
\subsection{No Rashba interaction}
\begin{figure}
\includegraphics[width=225pt]{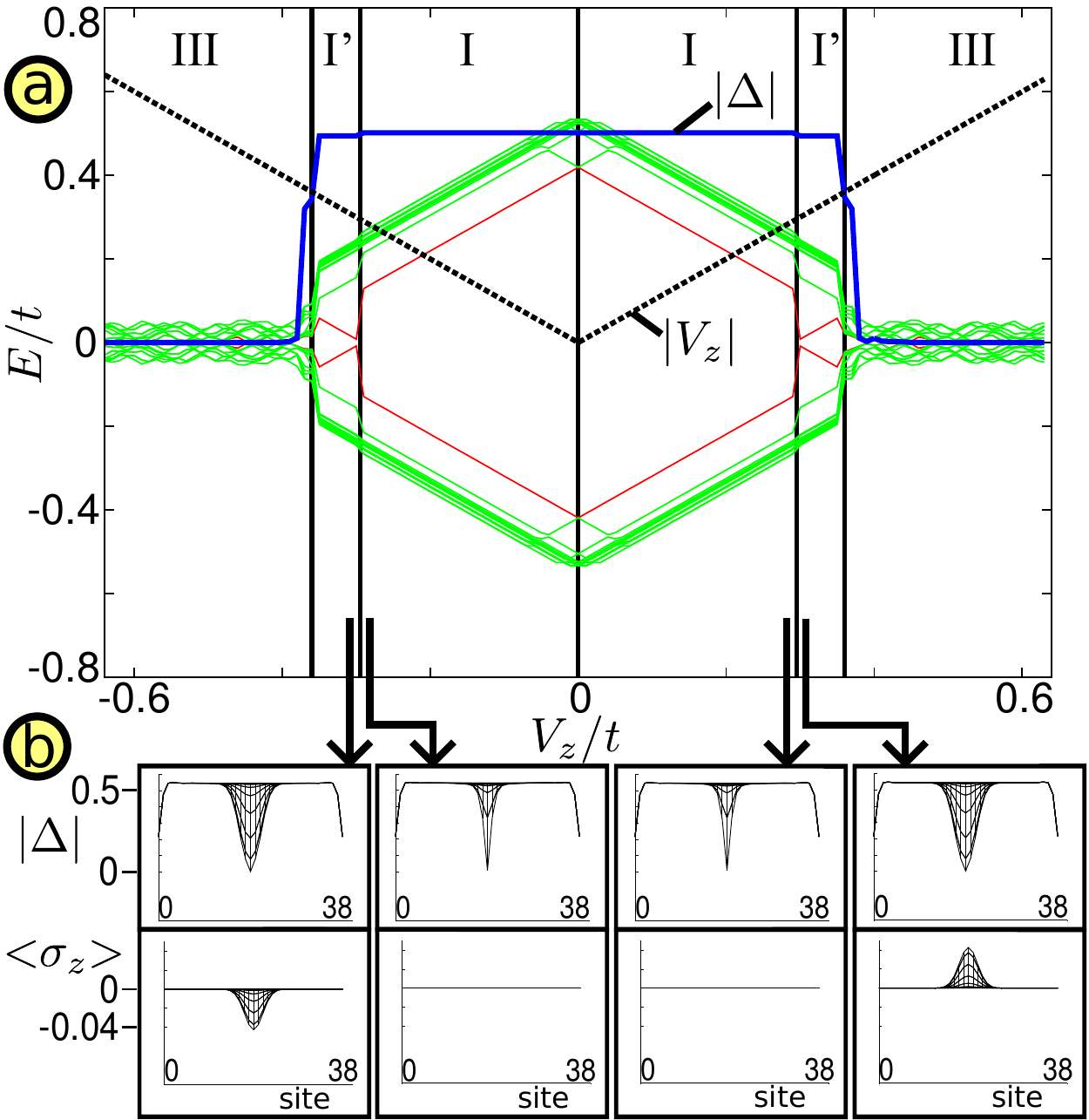}
\caption{(Color online) (a) Low-energy spectrum (thin lines), together with Zeeman splitting $|V_z|$ (black dotted line) and the average of the superconducting order parameter $|\Delta|$ (thick blue line) as a function of $V_z$. The two energy levels closest to zero are marked in red. A superconducting phase transition can be seen between the I' and III regions. (b) Real space profile of $|\Delta|$ and $\langle\sigma_z\rangle$. The profile changes continuously with $V_z$, except at the boundary between I and I' where it changes discontinuously as shown in the figures. Here $\alpha = 0$, $V_{sc} = 5.36$, $n = 1$.}
\label{Figure:No_Rashba}
\end{figure}

Before taking on the full problem we study the system in the absence of Rashba spin-orbit interaction, i.e.~for $\alpha = 0$. In Fig.~\ref{Figure:No_Rashba} we show the low-energy spectrum and the superconducting order parameter when $V_{sc} = 5.36$ and $n = 1$. A clear superconducting phase transition can be seen between the regions labeled I $\cup$ I' and III. Inside the superconducting phase I $\cup$ I' there is a sudden jump in the lowest energy levels and the appearance of a zero energy state marking the transition between I and I'. The energy level jump is correlated with an abrupt change in the superconducting order parameter in the vortex core, as well as an emergent magnetization of the core. In region I the vortex profile is narrow and there is no magnetization, while it is much wider in region I' where the core is also magnetized. 
There is thus a competition between $\Delta$ and $V_z$ in the core with a local transition taking place at the boundary between I and I'. This competition takes place throughout the sample and is ultimately responsible for the transition into the non-superconducting phase III. However, in the vortex core the Zeeman splitting becomes dominant earlier because here the superconducting state is suppressed by a large rotational component, which gives an additional kinetic energy contribution.
Apart from the boundary between I and I', the vortex core profile changes smoothly with $V_z$.
The changes of the vortex core at the transition between I and I' should be possible to detect experimentally using scanning probes.
%

\subsection{Including Rashba interaction}
\begin{figure}
\includegraphics[width=225pt]{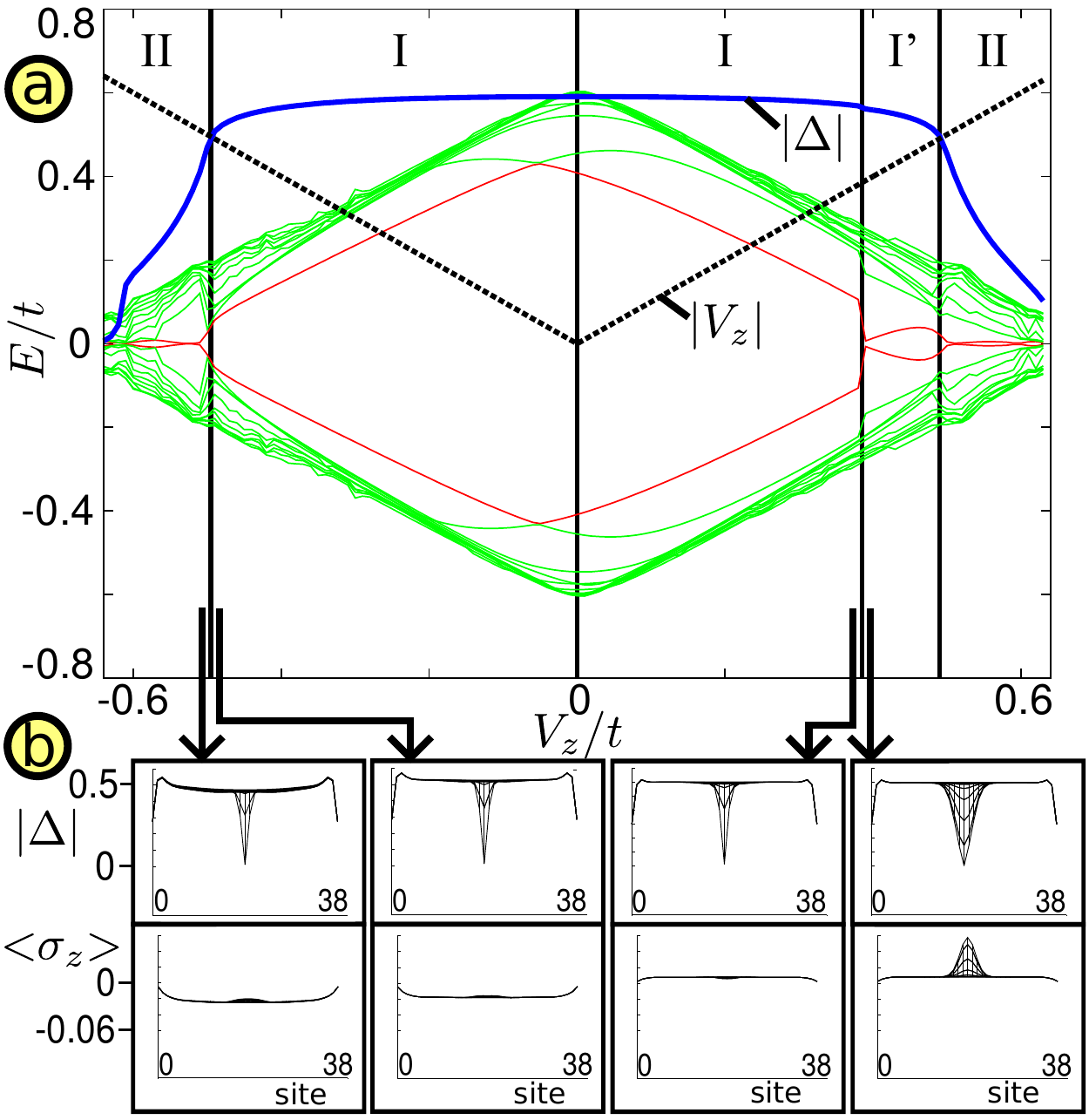}
\caption{(Color online) Similar to Fig.~\ref{Figure:No_Rashba} but for spin-orbit coupling $\alpha = 0.56$. The main differences are the appearance of a region labeled  II and the asymmetry with regards to $V_z$.}
\label{Figure:Alpha_0_28_scV_2_68_Vz}
\end{figure}

\begin{figure}
\includegraphics[width=225pt]{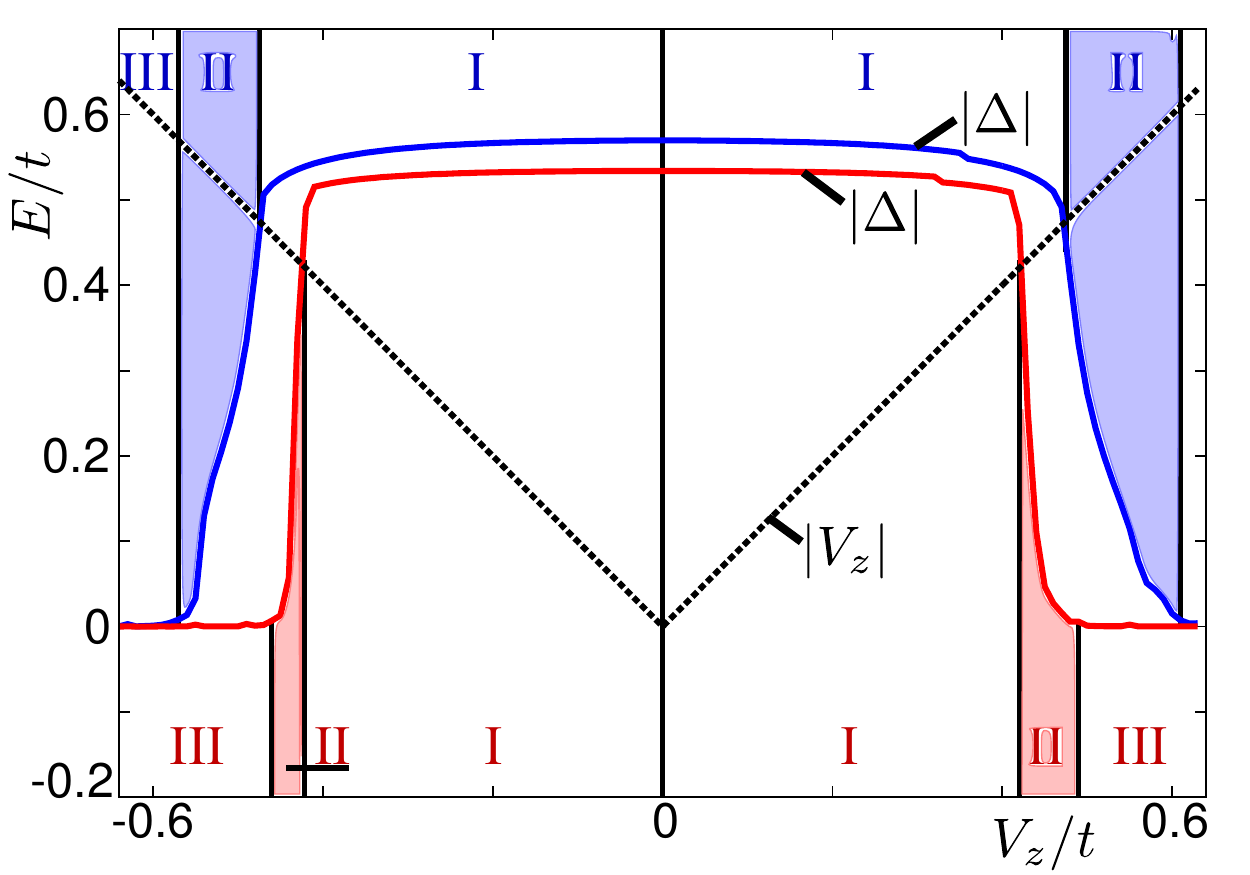}
\caption{(Color online) Zeeman splitting $|V_z|$ and average of the superconducting order parameter $|\Delta|$ for spin-orbit coupling $\alpha = 0.32$ (bottom red) and $\alpha = 0.48$ (top blue). Shaded areas represent the topologically non-trivial region II.}
\label{Figure:Orderparameter_alpha}
\end{figure}

We now continue to study the full Hamiltonian with $\alpha \neq 0$. Figure \ref{Figure:Alpha_0_28_scV_2_68_Vz} shows the low-energy spectrum and the superconducting order parameter for $\alpha = 0.56$, $V_{sc} = 5.36$, and $n = 1$. 
Two main effects are immediately visible when this result is compared to that of Fig.~\ref{Figure:No_Rashba}. First of all we see a clear asymmetry in both the energy spectrum and vortex structure as a function of $V_z$. Most notably we only find the I' region for positive $V_z$. We will discuss this asymmetry in depth in Sec.~\ref{Section:Sources_of_asymmetry}. The second observation is that the Rashba interaction has a large effect on the superconducting state, helping the superconducting order parameter to survive to higher Zeeman splitting. In addition, and for our purposes even more important, a finite $\alpha$ creates a smooth transition between the region with a large superconducting order parameter to that with no superconductivity. This means that a topologically non-trivial region II now appears, for which Eq.~(\ref{Equation:Topologically_non_trivial_phase_condition_2}) is satisfied. The lowest energy states in region II are Majorana fermions, as will be discussed in Sect \ref{Section:Majorana_fermions}. In Fig.~\ref{Figure:Orderparameter_alpha} we show a direct comparison between the superconducting order parameter for $\alpha = 0.32$ and $\alpha = 0.48$, which explicitly shows that a larger $\alpha$ creates a larger topologically non-trivial region II.

\begin{figure}
\includegraphics[width=225pt]{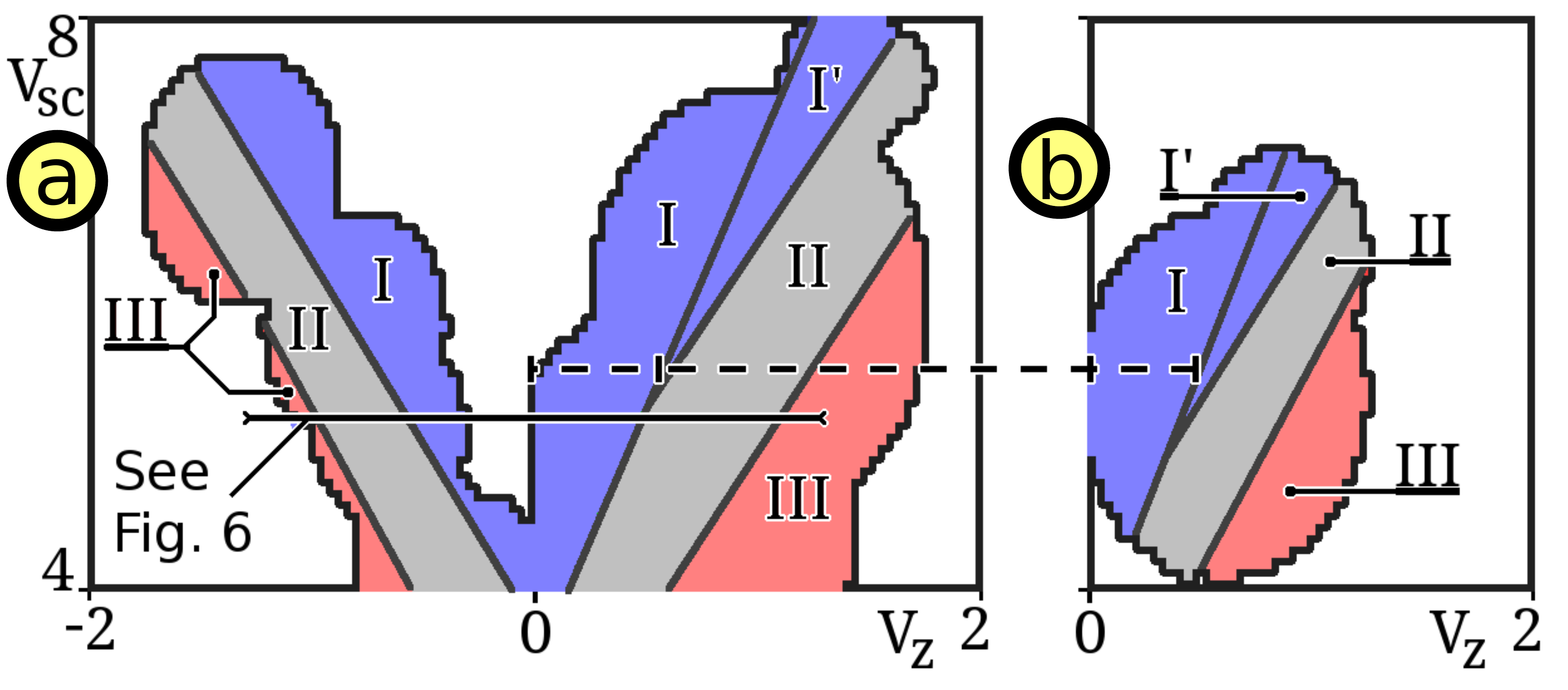}
\caption{(Color online) $(V_z,V_{sc})$ phase diagrams for (a) $\alpha = 1.04$ and (b) $\alpha = 0.8$, showing the occurrence of the four regions I, I', II and III. The white region has not been mapped. The lines are hand drawn approximations drawn on top of data obtained from calculations on a $26\times 26$ lattice. Solid line marks $V_z$ cut in Fig.~(\ref{Figure:No_I_prime_region}). Dashed line marks end point of I' wedge for $\alpha = 1.04$.}
\label{Figure:Phase_diagram}
\end{figure}

\begin{figure}
\includegraphics[width=225pt]{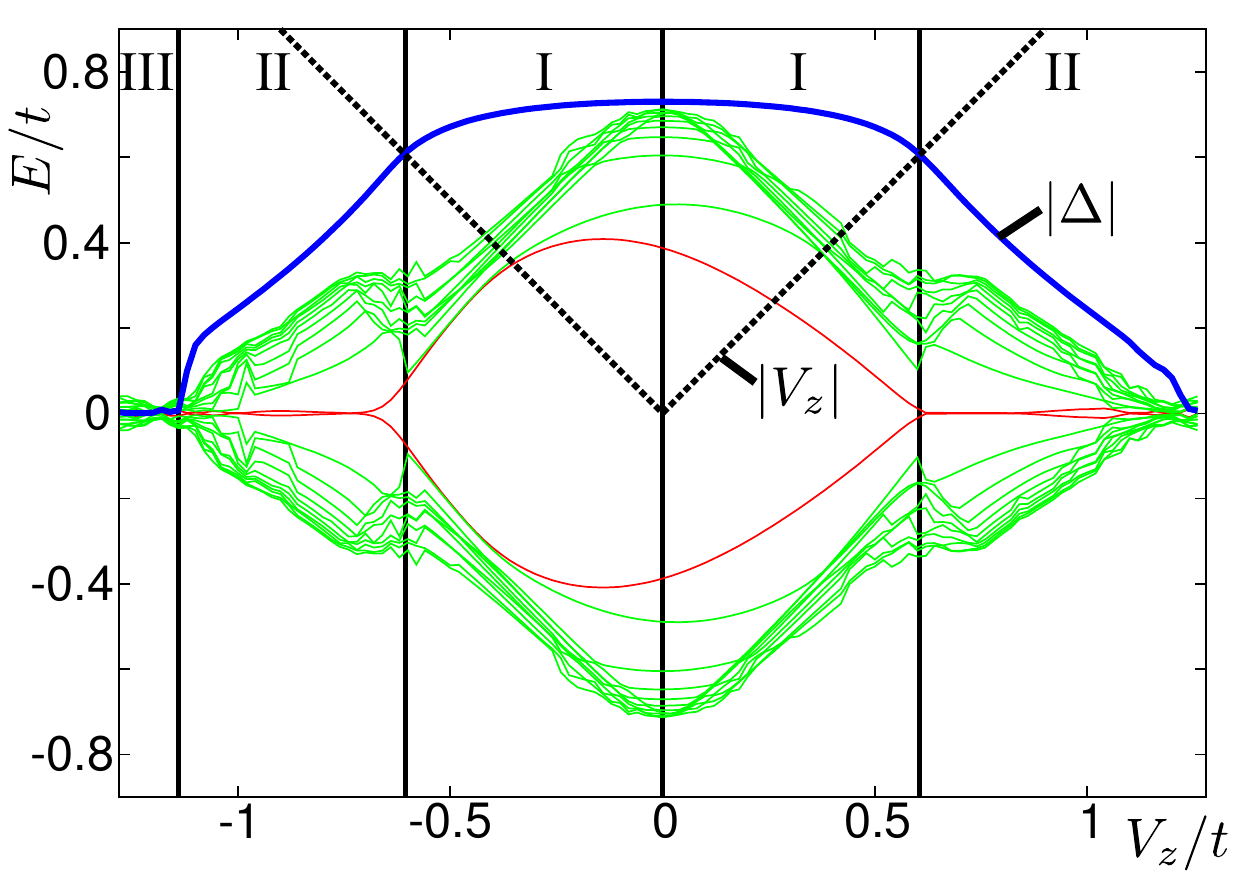}
\caption{(Color online) Similar to Fig.~\ref{Figure:No_Rashba} but for spin-orbit coupling $\alpha = 1.04$ (horizontal cut in Fig.~\ref{Figure:Phase_diagram}).}
\label{Figure:No_I_prime_region}
\end{figure}

In Fig.~\ref{Figure:Phase_diagram} the different phases I, I', II and III are visualized in $(V_z,V_{sc})$ phase diagrams. 
We once again see the occurrence of the I' region for positive $V_z$, but not for negative values. The I' region occurs as a wedge between I and II for positive $V_z$. In Fig.~\ref{Figure:No_I_prime_region} we plot the low-energy spectrum for the solid horizontal cut in Fig.~\ref{Figure:Phase_diagram} just below where the I' region disappears and we clearly see that the I' region is not present. 
On the right hand side of Fig.~\ref{Figure:Phase_diagram}, we display the $(V_z,V_{sc})$ phase diagram for a smaller $\alpha$, showing that the I' region now occurs at lower $V_{z}$ and $V_{sc}$. This is to be expected since the Rashba interaction helps superconductivity in the presence of a finite $V_z$. Thus, for decreasing Rashba coupling the vortex core gets magnetized, i.e.~enter region I', at lower values of $V_z$.

\subsection{Majorana fermions}
\label{Section:Majorana_fermions}
So far we have only discussed the topologically trivial and non-trivial regions, the superconducting phase transition and the vortex core magnetization. While interesting in themselves, our main reason for doing so was to be better equipped to identify which features are related to the existence of Majorana fermions, and which features are due to other effects. We now turn our attention to the Majorana fermions which have been predicted to occur inside the topologically non-trivial phase as zero energy modes in vortex cores.\cite{PhysRevB.77.220501,PhysRevLett.103.020401,PhysRevLett.104.040502,PhysRevB.82.134521} Majorana fermions, being essentially half of an ordinary fermion, necessarily comes in pairs, generally located in different vortex cores. In a system with only one vortex, the second Majorana fermion will instead be located on the edge of the system. In Fig.~\ref{Figure:Alpha_0_28_scV_2_68_Vz} we clearly see two near zero energy modes in the topologically non-trivial phase II. However, they show slightly oscillatory behavior, such that the mathematical criterion $E = 0$ for a Majorana fermion in a superconductor is not strictly satisfied, and there is also several other low-lying states close to zero in these regions.

To establish the Majorana fermion nature of the two lowest energy modes in region II we plot the spatial probability density for the $E \approxeq 0$ eigenstates inside region II. 
\begin{figure}
\includegraphics[width=225pt]{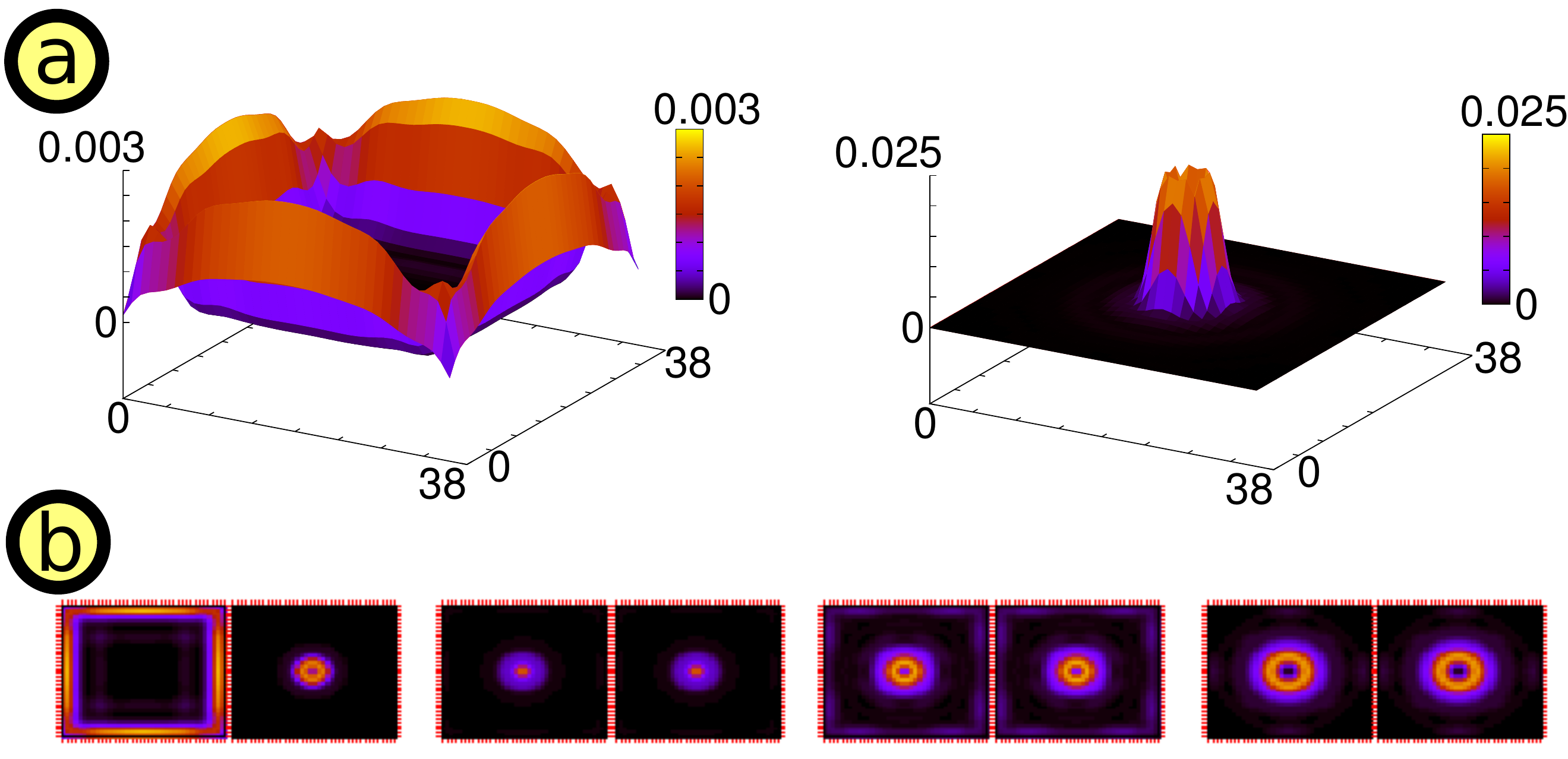}
\caption{(Color online) (a) Probability density for the two lowest energy states in region II on the left side of Fig.~\ref{Figure:Alpha_0_28_scV_2_68_Vz}, i.e.,~the Majorana states. (b) Left pair shows the same states from a top view, while the three next pairs are the probability densities for the next three pairs of states closest to zero energy. Note how only the Majorana fermion pair can be separated into one core and one edge state.}
\label{Figure:MajoranaFermions_alpha_0_28_scV_2_68_Vz}
\end{figure}
If plotted directly, these eigenstates appear as two states that have finite weight both at the edge and in the core. This is an artifact of the ambiguity in basis used to describe these (nearly) degenerate eigenstates. In Fig.~\ref{Figure:MajoranaFermions_alpha_0_28_scV_2_68_Vz} we therefore plot $\gamma_1 = \frac{1}{\sqrt{2}}\left(\gamma_{+} + \gamma{-}\right)$ and $\gamma_2 = \frac{1}{\sqrt{2}}\left(\gamma_{+} - \gamma_{-}\right)$ rather than $\gamma_+$ and $\gamma_-$, where $\gamma_+$ and $\gamma_-$ are the eigenstates from the numerical diagonalization. In the same figure we also plot the probability density for the next three pairs of states closest to $E = 0$. It is clear that the lowest-energy pair of states (leftmost figure) is very different from the other states. The two lowest energy eigenstates are spatially separable into one core and one edge state, clearly displaying their Majorana nature. Distinctly different, the next lowest eigenstates are non-separable vortex-centered states, thus combinable into one single normal electronic excitation. These are the equivalent of the Caroli-de Gennes-Matricon vortex states found in conventional $s$-wave superconductors.\cite{Caroli64}

Having established the Majorana fermion nature of the lowest-energy modes in region II we note that their non-zero energies are an artifact of a finite lattice. In fact, it has been shown that Majorana fermions that come close to each other experience interference effects.\cite{PhysRevLett.103.107001, PhysRevB.82.134521, NewJPhys.13.075009} We therefore draw the conclusion that the slight deviations from $E = 0$ are due to interference between the Majorana fermions in the core and on the edge.

\subsection{Zero mode at the vortex core transition}
Revisiting the vortex core transition that occurs between I and I' in Fig.~\ref{Figure:Alpha_0_28_scV_2_68_Vz}, we may now ask whether the zero mode that appears at this transition is also a Majorana fermion. An immediate objection to this would be that this is well inside the topologically trivial phase, where Majorana fermions have previously not been predicted to exist. However, the classification of the topological phases are done in a bulk calculation where the order parameter is constant, while vortex calculations inherently implies an order parameter with both varying phase and amplitude. We therefore push this question a little further.

\begin{figure}
\includegraphics[width=225pt]{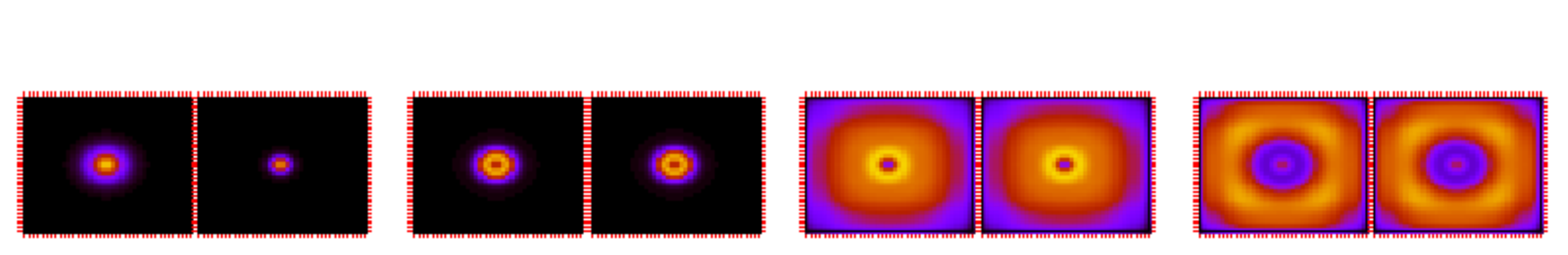}
\caption{(Color online) Same as Fig.~\ref{Figure:MajoranaFermions_alpha_0_28_scV_2_68_Vz} but immediately inside region I' on the right side of Fig.~\ref{Figure:Alpha_0_28_scV_2_68_Vz}. The lowest energy pair can not be separated into one core and one edge state and is not the type of separated Majoran pair described in \ref{Section:Majorana_fermions}, rather we interpret these as being Majoran fermions both trapped inside the vortex core. One at the center, and one at the boundary between the topologically non-trivial core and the topological trivial bulk.}
\label{Figure:LDOS_not_Majorana}
\end{figure}

In Fig.~\ref{Figure:LDOS_not_Majorana} we plot the probability density for the two lowest energy states just inside the I' region. It is clear that the first pair of states do not separate into one core and one sample edge state as they did in Fig.~\ref{Figure:MajoranaFermions_alpha_0_28_scV_2_68_Vz}, and we thus conclude that this zero mode is very different from the Majorana fermionic state observed when the bulk is in the topologically non-trivial phase II. We can, however, still view these states as having a Majorana origin by noting that the bulk is in the topologically trivial phase because $\Delta > V_z$, while closer to the core $\Delta$ becomes smaller and eventually passes through a point at which $\Delta < V_z$. In a bulk calculation this would mean a transition from the trivial to the non-trivial phase, and given that the vortex core is quite wide in the I' region, we interpret the results as the vortex core region being in a topologically non-trivial phase in region I'. Thus, we expect the I' region to host a central vortex core Majorana mode, as well as an edge Majorana mode, located at the boundary between the vortex core region (where $\Delta < V_z$) and the surrounding material which still is in the trivial I region.
However, due to to the small dimensions of the vortex core region, these two Majorana modes have a large spatial overlap causing significant interference effects and they will thus largely combine into a normal electron mode, as clearly seen in Fig.~\ref{Figure:LDOS_not_Majorana}. This explains the zero-energy state at the I to I' transition, as well as the finite energy of the lowest energy states inside region I'. Finally approaching the II region, the bulk gap closes in the full sample, facilitating the transport of the vortex core boundary Majorana mode to the true sample boundary, thus generating truly separated Majorana modes. Using this interpretation the I' phase can be viewed as a phase with a topologically trivial bulk, but with a non-trivial vortex core region.

The spatially overlapping Majorana fermions in region I' lack many of the properties that makes the Majorana fermions that comes in odd numbers in each core in phase II interesting. They can constitute an experimental challenge to the observation of the spatially separated Majorana fermions as they cause zero-energy signatures before entering the topologically non-trivial region II. However, in spite of these drawbacks these states can also be of experimental interest, especially as stepping stones towards finding the spatially separated Majorana fermions, described in Sec.~\ref{Section:Majorana_fermions}. The reason for this is clear if we consider Fig.~\ref{Figure:Alpha_0_28_scV_2_68_Vz}, which shows that these states are much more separated from the higher lying energy states than the Majorana fermions in phase II and could therefore be easier to detect. Furthermore, the transition into the I' region is more abrupt than that between I and II, which offers a potentially more unique experimental signature when tuning the Zeeman exchange coupling.
%
%
%

\subsection{Spectral asymmetries}
\label{Section:Sources_of_asymmetry}
\begin{figure*}
\subfigure{\includegraphics[width=235pt]{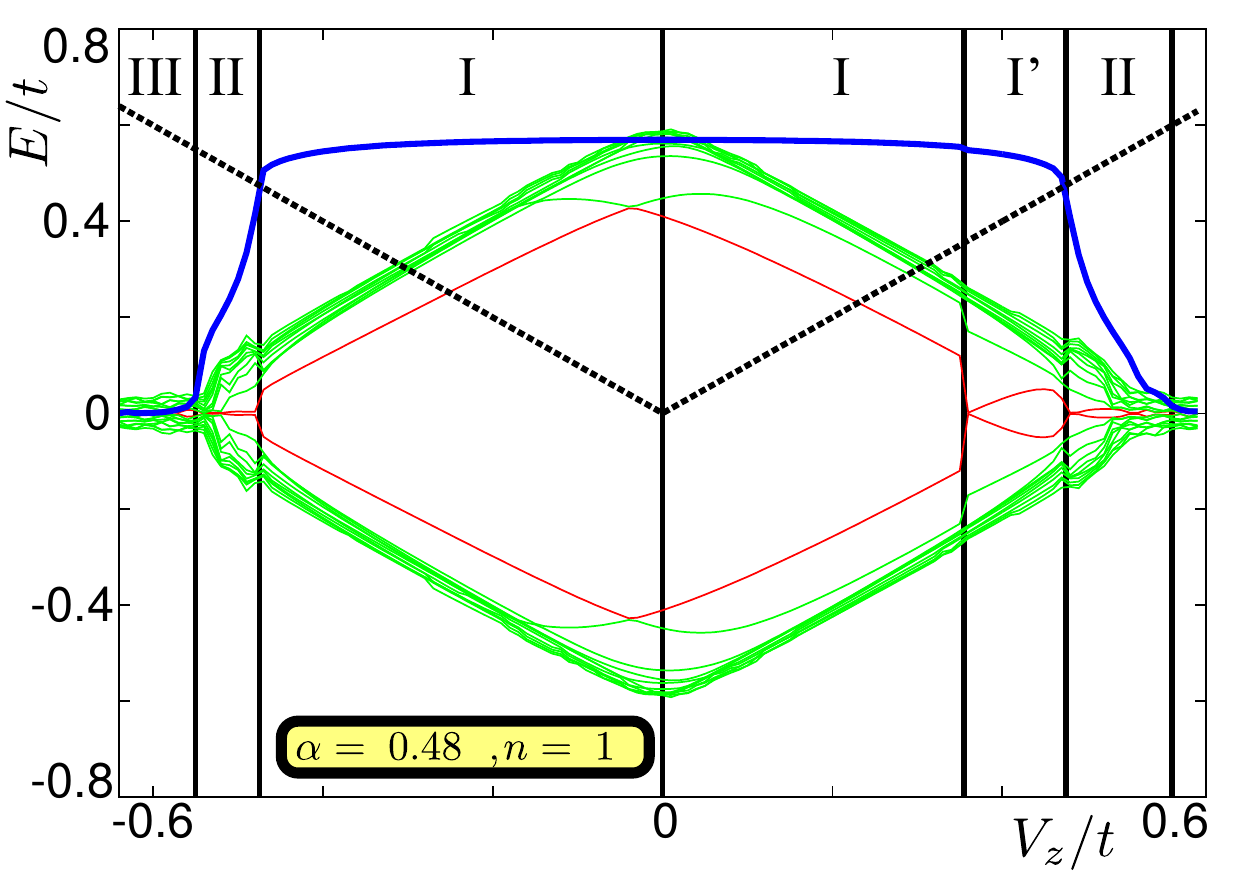}}
\subfigure{\includegraphics[width=235pt]{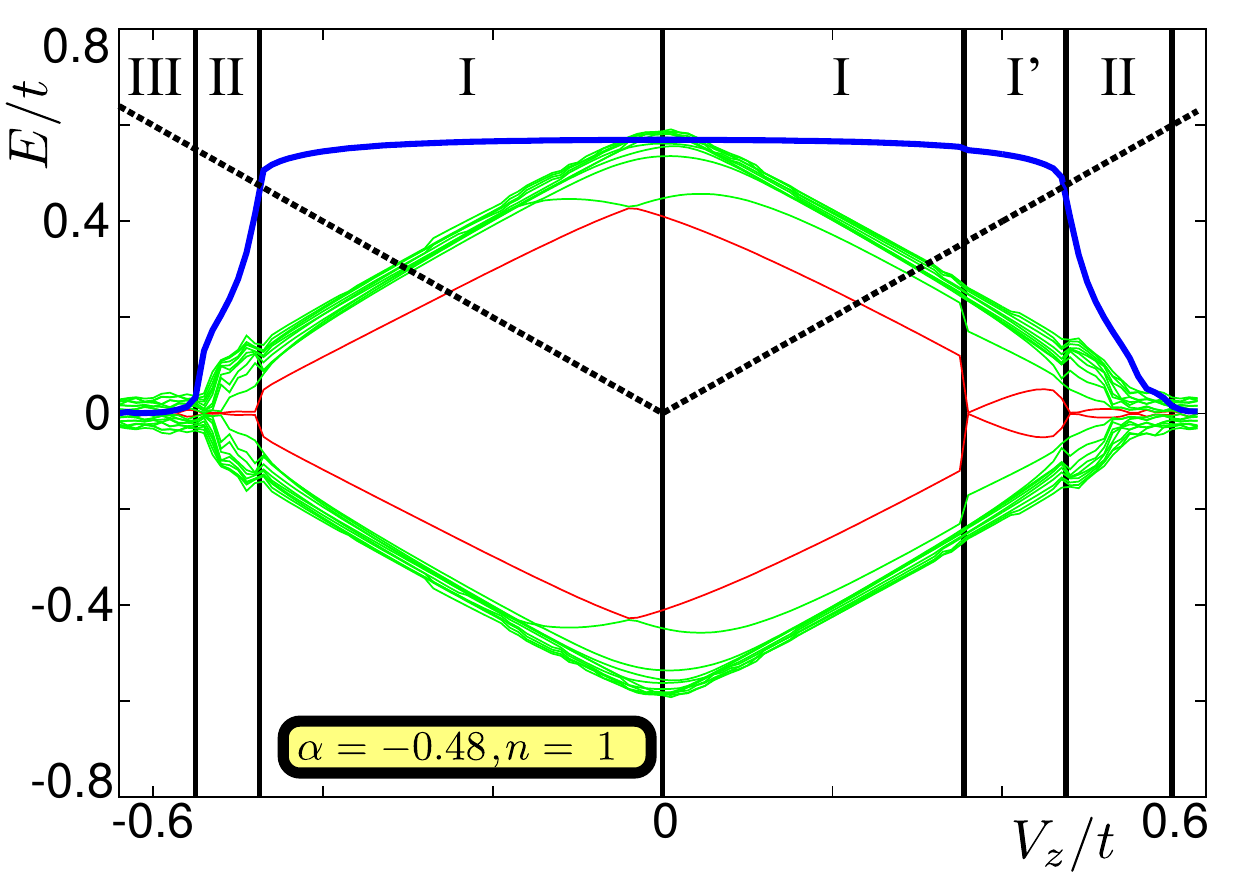}}\\
\subfigure{\includegraphics[width=235pt]{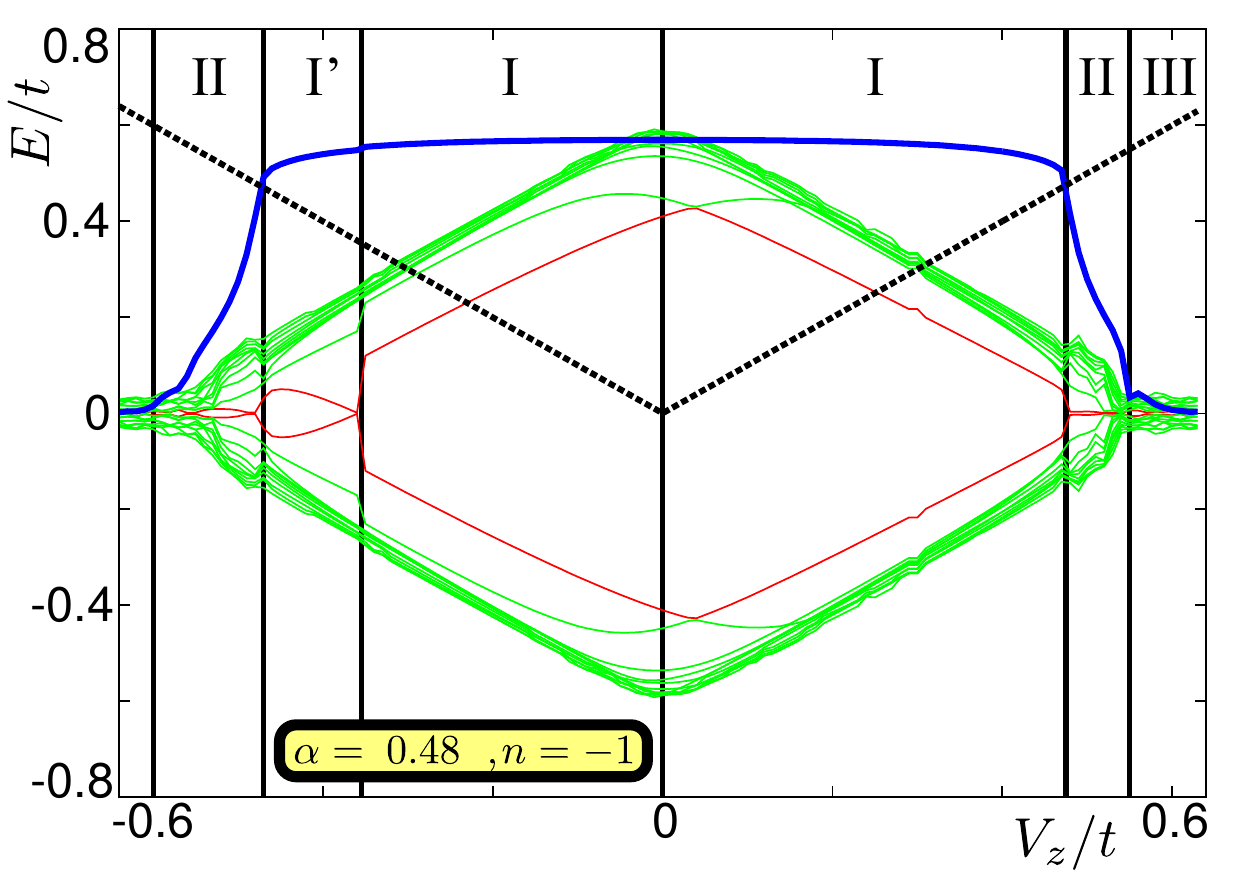}}
\subfigure{\includegraphics[width=235pt]{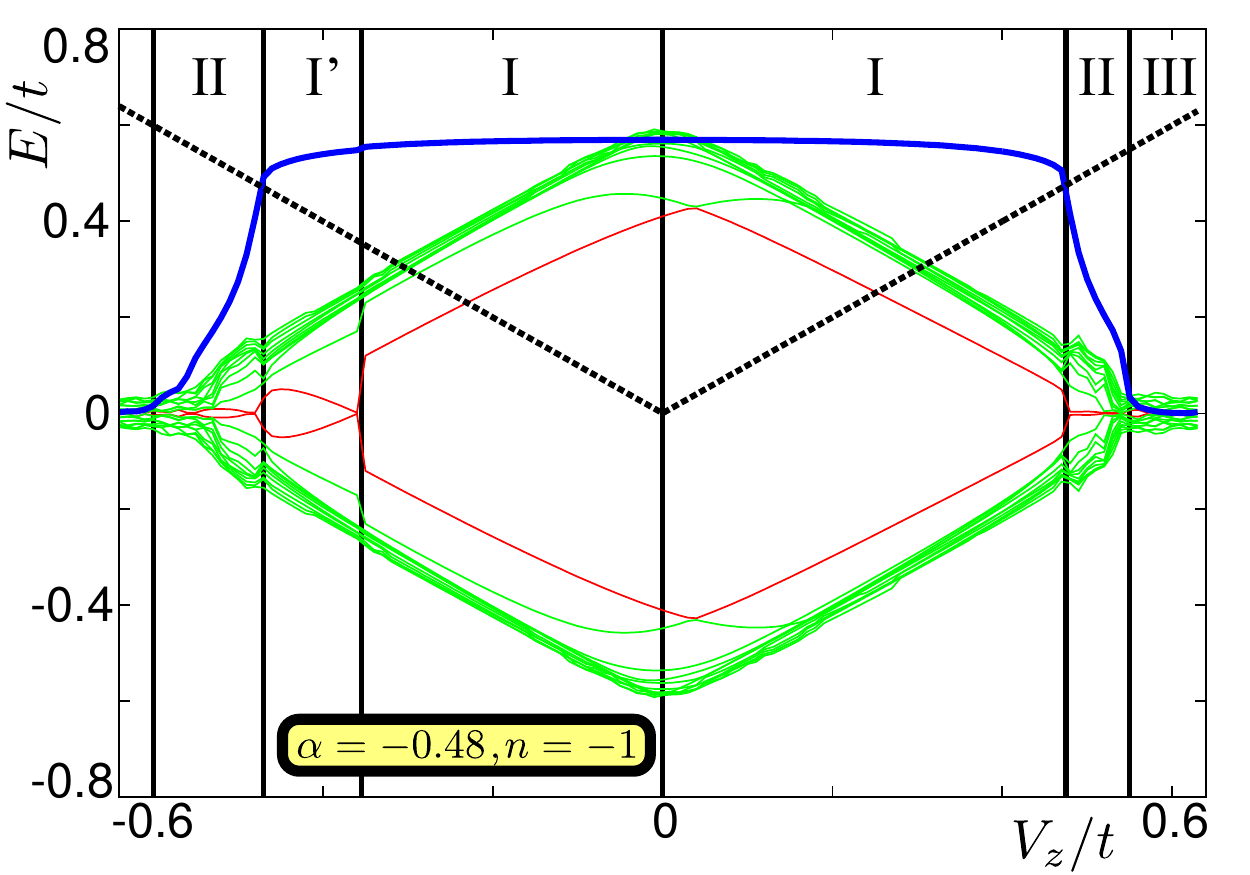}}\\
\caption{(Color online) Similar to Fig.~\ref{Figure:No_Rashba} but for spin-orbit coupling $|\alpha| = 0.48$ and different signs on $\alpha$ and vortex rotation $n$. The energy spectrum is asymmetric with respect to $V_z \rightarrow -V_z$ and $n \rightarrow -n$, but symmetric with respect to $\alpha \rightarrow -\alpha$ and $V_z, n \rightarrow -V_z, -n$.}
\label{Figure:EnergySpectraAsymmetries}
\end{figure*}

Finally, we address the asymmetry that is seen in the energy spectrum when $\alpha \neq 0$. It is clear from our numerical results that there is an asymmetry in the energy spectrum as well as vortex core order parameter and magnetization as function of $V_z$. In Fig.~\ref{Figure:EnergySpectraAsymmetries} the result of flipping the vortex direction $n \rightarrow -n$ and the sign of the spin-orbit interaction $\alpha \rightarrow -\alpha$ is shown for $|\alpha| = 0.48$. The energy spectrum is independent of the sign of the spin-orbit coupling, but the dependence of $V_z$ is flipped for reversed vortex rotation. The energy spectrum is thus asymmetric with respect to $V_z \rightarrow -V_z$ and $n \rightarrow -n$, but the results shows a symmetry with respect to the simultaneous parameter transformation $V_z, n \rightarrow -V_z, -n$.

In order to gain more understanding of the physical origin of these symmetries and asymmetries we consider the analytical continuum equivalent of Eq.~(\ref{Equation:Tight_binding_Hamiltonian}). The Hamiltonian, in the gauge where $\Delta$ has a constant phase, can then be written as\cite{PhysRevB.82.134521}
\begin{widetext}
\begin{equation}
\mathcal{H} = \left[\begin{array}{cccc}
D_1 + D_2 + D_3 & S_{+}^{(1)} + S_{-}^{(2)} & 0 & \Delta\\
S_{-}^{(1)} + S_{+}^{(2)} & D_1 + D_2 - D_3 & -\Delta & 0\\
0 & -\Delta & -D_1 + D_2 - D_3 & S_{-}^ {(1)} - S_{+}^{(2)}\\
\Delta & 0 & S_{+}^{(1)} - S_{-}^{(2)} & -D_1 + D_2 + D_3
\end{array}\right]
\label{Equation:Continuum_model}
\end{equation}
\end{widetext}
where
\begin{align}
D_1 &= -\frac{\hbar^2}{2m}\left(\frac{1}{r}\frac{\partial}{\partial r}\left(r\frac{\partial}{\partial r}\right) + \frac{1}{r^2}\frac{\partial^2}{\partial\theta^2}\right),\\
D_2 &= -\frac{i\hbar n}{r^2}\frac{\partial}{\partial\theta},\nonumber \\
D_3 &= -V_z, \nonumber \\
S_{\pm}^{(1)} &= -i\hbar\alpha
\left(\left(\sin(\theta) \pm i\cos(\theta)\right)\frac{\partial}{\partial r}\right. \nonumber \\
&\left. + \frac{1}{r}\left(\cos(\theta) \mp i\sin(\theta)\right)\frac{\partial}{\partial\theta}\right),\nonumber \\
S_{\pm}^{(2)} &= \frac{\alpha n}{2r}\left(\cos(\theta) \pm i\sin(\theta)\right) \nonumber.
\end{align}

Noting that the numerical calculations indicate that the system is symmetric for $\alpha = 0$, we start our analysis in this limit. The $S$ terms then disappears and it is easy to see that $V_z \rightarrow -V_z$ implies $D_3 \rightarrow -D_3$, while all other terms are unchanged. The effect on the Hamiltonian of this physical transformation, which corresponds to a reversal of the Zeeman field direction, can almost be undone by the similarity transformation $s H s^{-1}$, where
\begin{equation}
s = s^{-1} = \left[\begin{array}{cccc}
0 & 1 & 0 & 0\\
1 & 0 & 0 & 0\\
0 & 0 & 0 & 1\\
0 & 0 & 1 & 0
\end{array}\right].
\end{equation}
This transformation corresponds to a relabeling of up and down spins, or equivalently, a spin flip. The only place $sHs^{-1}$ differs from $H$ is a minus sign in front of the $\Delta$ term, which does not affect the energy spectrum. Physically, the Zeeman term breaks the symmetry along the $z$-axis, and when it is reversed it suffices for any eigenstate in the original system to reverse its spin to become an eigenstate with the same energy in the new system where the Zeeman field is reversed.
Continuing our analysis at $\alpha = 0$ we also look at the effect of $n \rightarrow -n$. This transformation, which corresponds to a reversal of the vortex rotation direction, gives $D_2 \rightarrow -D_2$. This physical transformation can be undone by the coordinate transformation $\theta \rightarrow -\theta$. This means that any eigenstate in the system with vortex rotation direction $n$ can be brought into an eigenstate with the same energy for a system with vortex rotation $-n$ by simply reversing the rotation direction for the state.
Slightly rephrasing the results for $\alpha = 0$, the Zeeman splitting and vortex rotation direction can each be seen to provide an independent $Z_2$ symmetry with respect to reversal in the $z$-direction, that is there exists a $Z_2\times Z_2$ symmetry. As a consequence, for each eigenstate after either of these transformations, a partner eigenstate with the same energy can be found by flipping the spin direction or rotation direction, respectively, for that state.

The situation becomes more complicated for finite $\alpha$. The transformation $V_z \rightarrow -V_z$ can no longer be undone by the similarity transformation $sHs^{-1}$, because the off-diagonal $S$ elements are interchanged during this process. Similarly, the physical transformation $n \rightarrow -n$ is no longer counteracted by the coordinate transformation $\theta \rightarrow -\theta$, because, again, the $S$ terms are modified in a non-trivial way in this process.
However, if we reverse both the magnetic field and the vortex direction, i.e.~$V_z, n \rightarrow -V_z, -n$, then the coordinate transformation $\theta \rightarrow -\theta$, followed by the similarity transformation $sHs^{-1}$ brings back the original $H$ apart from an additional minus sign on all off-diagonal terms, which does not influence the energy spectrum.

Mathematically, the spectral asymmetries in $V_z$ and $n$ are directly related to the introduction of the Rashba spin-orbit dependent $S$ terms in Eq.~(\ref{Equation:Continuum_model}). To understand this we note that when only the Rashba interaction is present, such that the only term that is non-zero is $S_{\pm}^{(1)}$, then the spin lies in-plane and is locked perpendicular to the direction of motion. This is evident from the Rashba spin-orbit interaction in Cartesian coordinates: $\alpha(\mathbf{k}\times\boldsymbol{\sigma})_z$. It is also clear that the energy depends on which of the two directions perpendicular to the direction of motion the spin locks onto, and that the energy is preserved under a simultaneous reversal of the directions of motion and spin.
Coming back to the full problem, the Zeeman term will cause an out-of-plane tilting of the spin. 
Taking also into account the kinetic and superconducting terms, the exact form of the eigenstates can be expected to be quite complicated. However, for our purpose it is not important to know the exact eigenstates of the Hamiltonian, rather it suffices to draw the following conclusions about how three of the terms influence the eigenstates:
\begin{itemize}
\item{$D_3$:}  Each eigenstate gets an energy contribution due to the $V_z$ term, which depends on the spin being tilted up- or downwards.
\item{$S_{\pm}^{(1)}$ ($S_{\pm}^ {(2)}$):} Each eigenstate gets an energy contribution from the Rashba spin-orbit interaction, which depends on the relative orientation of the direction of motion and the in-plane spin component.
\item{$D_2$ ($S_{\pm}^ {(2)}$):} Each eigenstate gets an energy contribution from the superconducting vortex term, which depends on the particle motion being with or against the rotation of the vortex.
\end{itemize}
Above $S_{\pm}^{(2)}$ appears in parentheses because it contributes to a mixture of the two latter effects. With regards to the spin-flip operation, the $\pm$-index transforms in the same way as for the $S_{\pm}^{(1)}$ term. However, under the $n \rightarrow -n$ and $\theta \rightarrow -\theta$ transformations, $S_{\pm}^{(2)}$ is multiplied by a negative sign relative to the $S_{\pm}^{(1)}$ term.

\begin{figure}
\includegraphics[width=225pt]{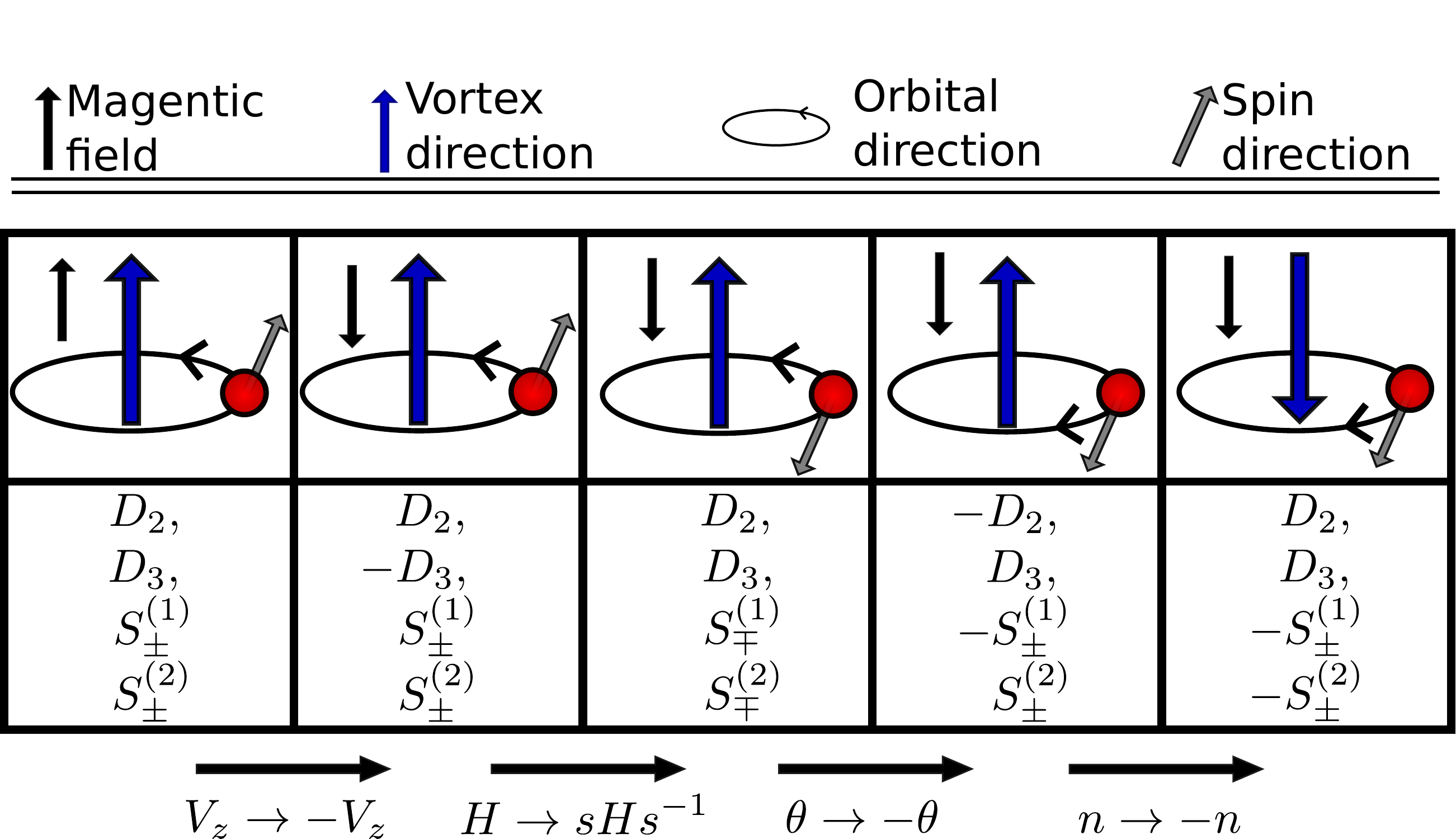}
\caption{(Color online) Schematic figure depicting the effect of applying parameter and coordinate transformations to the Hamiltonian in Eq. (\ref{Equation:Continuum_model}). Only the relative sign between the $S$ terms is important for the energy spectrum: $\det(H - IE) = 0$, and the first and last stage is therefore energetically equivalent. The three middle stages are always energetically equivalent as they only differ by a (non-physical) coordinate transformation.
}
\label{Figure:Asymmetries_schematic}
\end{figure}

With this we are finally able to explain the results in Fig.~\ref{Figure:EnergySpectraAsymmetries}. We begin with the observation that for $V_z \rightarrow -V_z$ the off-diagonal $S$ terms obstructs attempts to get back eigenstates with the same energy through a spin flip. The reason is that the Rashba term does not in general contribute the same amount of energy to the eigenstates after this transformation, unless the rotation direction for the states are reversed at the same time, which requires the coordinate transformation $\theta \rightarrow -\theta$. However, the energy contribution from having a particle rotating either clockwise or counter-clockwise around the vortex is not the same unless the vortex rotation direction is also changed. Therefore the energy spectrum is preserved only under a simultaneous reversal of $V_z$ and $n$.
A schematic depiction of this analysis is given in Fig.~\ref{Figure:Asymmetries_schematic}.
We see that the asymmetry in the energy spectrum is the combined effect of Zeeman, Rashba, and vortex rotation. The $n \rightarrow -n$ asymmetry can be worked out following a completely analogues procedure and can be visualized by reading the steps in Fig.~\ref{Figure:Asymmetries_schematic} in reversed order.

Although the description above provides insight into how the asymmetry in $V_z$ and $n$ arises at a microscopic level, it is also useful to rephrase the results in terms of the $Z_2\times Z_2$ symmetry introduced above for the case $\alpha = 0$. Remembering that this symmetry arises because the spin and angular momentum are independent degrees of freedom for $\alpha = 0$, which couple to the Zeeman splitting and vortex rotation direction, respectively. Once a finite spin-orbit interaction is introduced these two degrees of freedom are mixed and splits the $Z_2\times Z_2$ symmetry into two distinct configurations, each having a $Z_2$ symmetry. The two distinct configurations are those for which the Zeeman splitting and vortex rotation direction are either parallel or anti-parallel, each of which can be achieved in two different, but energetically, equivalent ways.

Finally, we note that the symmetry observed under the reversal of $\alpha$ is very similar to that expected from a global phase transformation in $\Delta$. This can be seen by concluding that $\alpha \rightarrow -\alpha$ has the effect of transforming $S_{\pm}^{(1)}, S_{\pm}^{(2)} \rightarrow -S_{\pm}^{(1)}, -S_{\pm}^{(2)}$. Such a transformation leaves $\det(H - IE)$ invariant and the energy spectrum is therefore guaranteed to be preserved. 
In fact, it can be shown that the energy spectrum is invariant under global $U(1)$ phase transformations of both $\alpha$ and $\Delta$, and that these can be performed independently of each other such that these parameters gives rise to a $U(1)\times U(1)$ symmetry. This symmetry can be traced back to the rotational $SO(2)$ symmetry around the $z$-axis passing through the vortex core. Even though this symmetry is broken down into a $C_4$ symmetry for the square lattice itself, the full $SO(2)$-symmetry is inherited independently by both the $\alpha$ and $\Delta$ parameters. From this point of view we can indeed expect a $SO(2)\times SO(2) \cong U(1)\times U(1)$ symmetry in these parameters.


\section{Conclusions}
We have studied a ferromagnetic 2D Rashba spin-orbit coupled semiconductor with proximity-induced spin-singlet \textit{s}-wave superconductivity by using the self-consistent Bogoliubov-de Gennes method on a square lattice. The calculations confirm the existence of Majorana fermions in vortex cores in the topologically non-trivial phase. Strong Rashba spin-orbit interaction makes the non-trivial phase larger as it counteracts the pair-breaking effects of the Zeeman spin splitting on the superconducting state.
We have also established the existence of a second superconducting phase transition within the topologically trivial phase, characterized by a finite vortex core magnetization and a wider vortex profile. A zero-energy mode appears at the transition between these two superconducting states. This demonstrates the existence of zero-energy excitations in the vortex core other than the spatially well separated Majorana fermions which appear inside the topologically non-trivial phase and they can provide both experimental obstacles as well as opportunities. We have also found a pronounced asymmetry in the vortex core energy spectrum with respect to the Zeeman splitting and the vortex rotation direction. This asymmetry is a consequence of the interaction between the, partially competing, Zeeman, Rashba, and vortex rotation terms in the Hamiltonian, resulting in energy states with a delicate dependence on the spin and orbital coordinates. Together these results show on a complex behavior of the vortex core energy spectrum in Rashba spin-orbit coupled superconductors. 

\acknowledgements
We are grateful to A.~V.~Balatsky, M.~Fogelstr\"om, T.~H.~Hansson, and V.~Lahtinen for discussions and the Swedish research council (VR) for financial support. The computations were performed on resources provided by SNIC through Uppsala Multidisciplinary Center for Advanced Computational Science (UPPMAX) under Project s00112-187.

\end{document}